# Spectrally-selective dynamic radiative thermoregulation via phase engineering

Qizhang Li[1]†, Yuanke Chen[1]†, Zhuang Luo[2], Chenxi Sui[1], Xubing Wu[1], Ronghui Wu[3], Qingsong Fan[1],

Ching-Tai Fu[1], Pei-Jan Hung[1], Gangbin Yan[4], Genesis Higueros[1], Ting-Hsuan Chen[1], Po-Chun Hsu[1]*

[1]*Pritzker School of Molecular Engineering, University of Chicago, Chicago, IL 60637, USA*

[2]*Computer Science Department, Worcester Polytechnic Institute, Worcester, MA 01609, USA*

[3]*School of Materials Science and Engineering, Nanyang Technological University, Singapore, 639798, Singapore*

[4]*Department of Physics, Stanford University, Stanford, CA 94305, USA*

†These authors contributed equally to this work

*Corresponding email: pochunhsu@uchicago.edu

**Abstract**

Maintaining comfortable temperatures for buildings, humans, and devices consumes a substantial portion of global energy, underscoring the urgent need for energy-efficient thermoregulation technologies. Dynamic radiative thermal emitters that can switch between passive cooling and heating modes offer a promising solution, but most existing devices exhibit broadband optical responses, resulting in unwanted parasitic heat exchange and limited performance. Here, we introduce an elegant strategy that uses a dielectric cap to transform broadband metal-insulator transition (MIT) materials into spectrally selective dynamic emitters. This design creates a highly tunable Fabry–Pérot cavity, enabling a tailored thermal emission spectrum by engineering the reflected-wave phase profile. Our Fresnel-formalism-based phasor diagram analysis reveals two key routes for realizing high spectral selectivity: a high-index dielectric cap and a low-loss metallic MIT state, which are further validated by Bayesian optimization. Following this principle, we demonstrated a wide-angle spectrally-selective thermoregulator operating in the atmospheric transparency window (8–13 μm), where the thermal emittance can be electrically tuned from about 0.2 to 0.9 through reversible copper electrodeposition on a germanium cavity. Furthermore, this strategy can be extended to multispectral electrochromic windows, enabling switching between solar heating and spectrally-selective radiative cooling. Our work establishes a versatile and generalizable paradigm for spectral engineering of dynamic thermal emitters, opening opportunities in energy-efficient buildings, wearable thermal comfort, spacecraft thermoregulation, and multispectral camouflage.

**Main**

The accelerated rise in global temperatures is driving more frequent and intense heat events worldwide, exacerbating health risks and increasing energy consumption (*1, 2*). In densely populated cities, urban infrastructures such as buildings and paved surfaces absorb and retain large amounts of solar radiation, resulting in the urban heat island effect, where surface temperatures can exceed 70 °C (*3*). This not only elevates the risk of heat-related illness and mortality but also substantially increases the cooling demand, reinforcing the vicious cycle of energy use and greenhouse gas emissions. Passive radiative cooling has emerged as a promising energy-efficient strategy to break this cycle, by dissipating heat directly to the cold universe through the atmospheric transparency window (ATW) (*4–7*). Over the past decades, radiative cooling has demonstrated significant energy-saving potential across diverse scenarios (*8–12*), progressing from laboratory demonstrations toward real-world applications (*13–15*). However, the pronounced heat island effect raises a major practical challenge to this technique: most radiative coolers are installed vertically and face hot surroundings, particularly in buildings and wearable applications (*16–19*). This results in unwanted radiative heat gain from the hot surroundings, severely suppressing their cooling performance. A promising pathway to address this issue is developing spectrally selective thermal emitters with dominant emission in the ATW wavelength range of 8–13 μm, which have showcased considerable improvement in cooling performance compared to the broadband counterparts (*18*, *19*).

Another practical challenge for radiative cooling technologies arises from fluctuating temperature conditions across seasons, compounded by the increasing frequency of extreme weather events such as heat waves and cold spells (*20*). In these cases, radiative coolers can overcool objects and increase heating costs. To cope with such fluctuations, dynamic radiative thermoregulators with switchable cooling and heating modes are essential (*21*). The key to achieving these devices lies in active materials with tunable optical properties, particularly in the mid-infrared (mid-IR) range. To date, extensive progress has been made using phase change materials (*22*, *23*), metal oxides (*24*–*26*), semiconductors (*27*, *28*), conjugated polymers (*29*, *30*), and reversible metal electrodeposition (RME) systems (*31*–*33*). However, these materials typically exhibit broadband optical responses, which greatly limits their ability to deliver desired spectral selectivity for efficient radiative control in the ATW (Table S1). Achieving selectivity in thermoregulators has often relied on integrating active materials with elaborate photonic structures such as multilayer stacks or lithographic patterning (*23*, *34*), which complicate fabrication, restrict material compatibility, and hinder large-scale applications. For example, in electrochromic systems, active materials must be placed at the bottom of photonic structures due to the highly absorptive electrolytes (*25*, *30*, *35*), which can compromise or even preclude their integration with complex optical designs. Consequently, there remains a critical need for a simple yet effective strategy to create dynamic thermal emitters with tailored spectral responses to meet the growing demand for efficient thermoregulation.

Here, we demonstrate a versatile approach to realize dynamic thermal emitters with desirable spectral selectivity and tunability by simply introducing a dielectric cap on a metal-insulator transition (MIT) material. Our phasor diagram analysis reveals that high spectral selectivity requires a high-index dielectric cap and a low-loss metallic state of the MIT material, which is further validated by Bayesian optimization. With this principle, we experimentally realized a spectrally selective dynamic thermal emitter operating in the ATW, achieving an average emissivity contrast of 0.57 and a peak modulation of 0.76 at ~10 μm through the RME of copper (Cu) on a germanium (Ge) cavity. We further extended this strategy to an electrochromic window by performing RME on a transparent high-index dielectric composed of a polymer matrix embedded with indium tin oxide nanocrystals (ITO NCs). This window can simultaneously block solar radiation and enable spectrally selective mid-IR emission in the cooling mode, while transmitting sunlight and suppressing radiative heat loss for heating purposes. Overall, this approach opens new possibilities for dynamic spectral control in emerging thermoregulation applications.

**Results**

**Strategy to realize spectrally-selective dynamic radiative thermoregulators**

Active materials form the basis of dynamic thermoregulators, and phase-change materials, especially MIT materials, are particularly appealing for their gigantic tunability of optical properties (*36*). Here, we generalize our discussion and broadly define MIT materials as those that can be switched between metallic and dielectric states, regardless of their physical forms, switching stimuli, or working principles. For a bulk MIT material, the metallic state typically exhibits high reflectivity ($R$) across the entire mid-IR range and consequently low thermal emissivity ($\varepsilon$), according to Kirchhoff's law ($\varepsilon = 1 - R$) under thermal equilibrium. This high reflectivity originates from the strong free-carrier oscillations at wavelengths longer than the plasma wavelength, which generally falls in the visible to ultraviolet region for many MIT materials (*37*). In contrast, the insulating state features broadband low reflectivity and thus strong thermal emission (or transmission, if the material has few or weak molecular vibrational modes within the thermal radiation wavelength range). Through this phase transition, MIT materials can achieve substantial emissivity modulation; however, the spectral selectivity is inherently limited (left panel of Fig. 1a and Fig. 1b).

To realize a spectrally selective dynamic thermoregulator (middle and right panels of Fig. 1a), our idea is to enhance the spectral selectivity of MIT materials, without compromising their tunability, by introducing a dielectric cap. This simple configuration makes the approach broadly compatible with diverse MIT materials, including solid-state materials such as vanadium dioxide ($VO_2$) (*38, 39*) and electrochemical platforms with liquid or gel electrolytes like conjugated polymers and RME systems (*30, 40, 41*). In the metallic state, the dielectric cap forms an asymmetric Fabry-Pérot (FP) cavity, enabling spectrally selective thermal emission via FP resonance. To better understand its optical behavior, we describe the reflectivity as the sum of the direct Fresnel reflection coefficients of the top surface ($r_{\text{dir}}$) and the FP reflection ($r_{\text{FP}}$) which accounts for multiple internal reflections within the cavity (Fig. 1c; see Supplementary Text 1 for details) (*38, 42*). Accordingly, the emissivity of this structure can be expressed as (*42*):

$$\varepsilon = 1 - |r_{\text{dir}} + r_{\text{FP}}|^2. \tag{1}$$

The emissivity reaches its maximum at the wavelength of destructive interference when $r_{\text{dir}}$ and $r_{\text{FP}}$ have comparable amplitudes and a π-phase difference (Figs. 1d and S1) (*42*). This design provides ease of spectral engineering as the resonant wavelength ($\lambda_0$) can be readily tuned by varying the dielectric cap thickness to accommodate different application scenarios. For instance, a peak emissivity around 10 μm is desirable for radiative cooling of objects near room temperature. More importantly, this configuration allows the tailoring of spectral selectivity by manipulating the phase difference $\Delta\varphi$ between $r_{\text{dir}}e^{i\pi}$ and $r_{\text{FP}}$ at an off-resonant wavelength ($\lambda_{\text{off}}$) (Fig. 1d). $\lambda_{\text{off}}$ is associated with the application's need for spectral range, e.g., 8–13 μm of the ATW. Note that here we introduce a π phase shift for $r_{\text{dir}}$ so that the destructive interference condition becomes $\Delta\varphi = 0$. Since the phase of $r_{\text{dir}}$ ($\approx \pi$) remains nearly constant, increasing the dispersion of the phase of $r_{\text{FP}}$ ($\varphi_{\text{FP}}$) makes the destructive interference decay more steeply with wavelength, leading to a sharper emission peak and enhanced spectral selectivity (Fig. 1d).

With a focus on $\varphi_{\text{FP}}$, we next explore the key factors governing the system's spectral selectivity. Based on the Fresnel equations, $r_{\text{FP}}$ can be written as:

$$r_{\text{FP}} = \frac{t_{12}t_{21}}{r_{12} + r_{23}^{-1}e^{-2i\varphi_2}}, \tag{2}$$

where $r_{ij}$ and $t_{ij}$ denote the reflection and transmission coefficients at the interface between media *i* and *j*, with 1, 2, and 3 representing air, dielectric cap, and MIT material, respectively (Fig. 1c); $\varphi_2$ is the accumulated phase during light propagation through the dielectric cap. Equation (2) suggests that the magnitude of $\varphi_{\text{FP}}$ ($|\varphi_{\text{FP}}|$) is determined by the vector sum of $r_{12}$ and $r_{23}^{-1}e^{-2i\varphi_2}$, denoted as $r_{\text{I}}$ and $r_{\text{II}}$, respectively. This is illustrated by the phasor diagram in the inset of Fig. 1d where the phase angles of $r_\mathrm{I}$ and $r_\mathrm{II}$ are fixed at a given wavelength $\lambda$ ($\varphi_\mathrm{I} \approx \pi$ and $\varphi_\mathrm{II} \approx \pi\left(1 - \frac{\lambda_0}{\lambda}\right)$, see Supplementary Text 2 for details). Accordingly, the phasor diagram reveals two distinct routes to increase $|\varphi_\mathrm{FP}|$ at $\lambda_\mathrm{off}$ for achieving higher spectral selectivity: (i) increasing the magnitude of $r_\mathrm{I}$ (left panel of Fig. 1e), or (ii) decreasing the magnitude of $r_\mathrm{II}$ (left panel of Fig. 1f). Both routes correspond to enlarging the magnitude of Fresnel reflection coefficients of the top and bottom interfaces of the dielectric cap (Fig. S2), as explicitly verified in Supplementary Text 2.

For the top interface of the dielectric cap, increasing its reflection magnitude necessitates a dielectric with a higher refractive index $n$ (Fig. 1e). This trend is confirmed by the calculated phasor diagrams with different dielectric-cap indices (Fig. S3). The corresponding spectral-phase calculations (Fig. S4) show that a higher index causes $\varphi_\mathrm{FP}$ to deviate more significantly from the destructive interference condition ($\Delta\varphi = 0$) as the wavelength moves away from $\lambda_0$, thereby enhancing the spectral selectivity (right panel of Fig. 1e; Fig. S5). Furthermore, the high index largely suppresses the angular dispersion of the resonant wavelength, maintaining hemispherical emission within the target spectral range, which is especially important for practical radiative cooling applications (see Supplementary Text 3). It is noteworthy that a high-index dielectric cap is also essential for achieving large emissivity contrast during the MIT switching, as it ensures high reflectivity (i.e., low emissivity) in the insulating state. For the bottom interface of the dielectric cap, the reflection magnitude is primarily determined by the MIT material in its metallic state (Fig. 1f). In this case, spectral selectivity is governed by the metallic-state loss of the MIT material. Lower metallic loss enlarges the reflection magnitude of the bottom interface and thereby increases $|\varphi_\mathrm{FP}|$ at $\lambda_\mathrm{off}$ (Figs. S6 and S7), resulting in a sharper emissivity peak and improved spectral selectivity (right panel of Fig. 1f; Fig. S8). Taken together, these results establish a general and rigorous design principle for dynamic thermal emitters: combining a high-index dielectric cap with an MIT material possessing a low-loss metallic state enables simultaneous realization of high spectral selectivity and tunability, as illustrated in Fig. 1g.

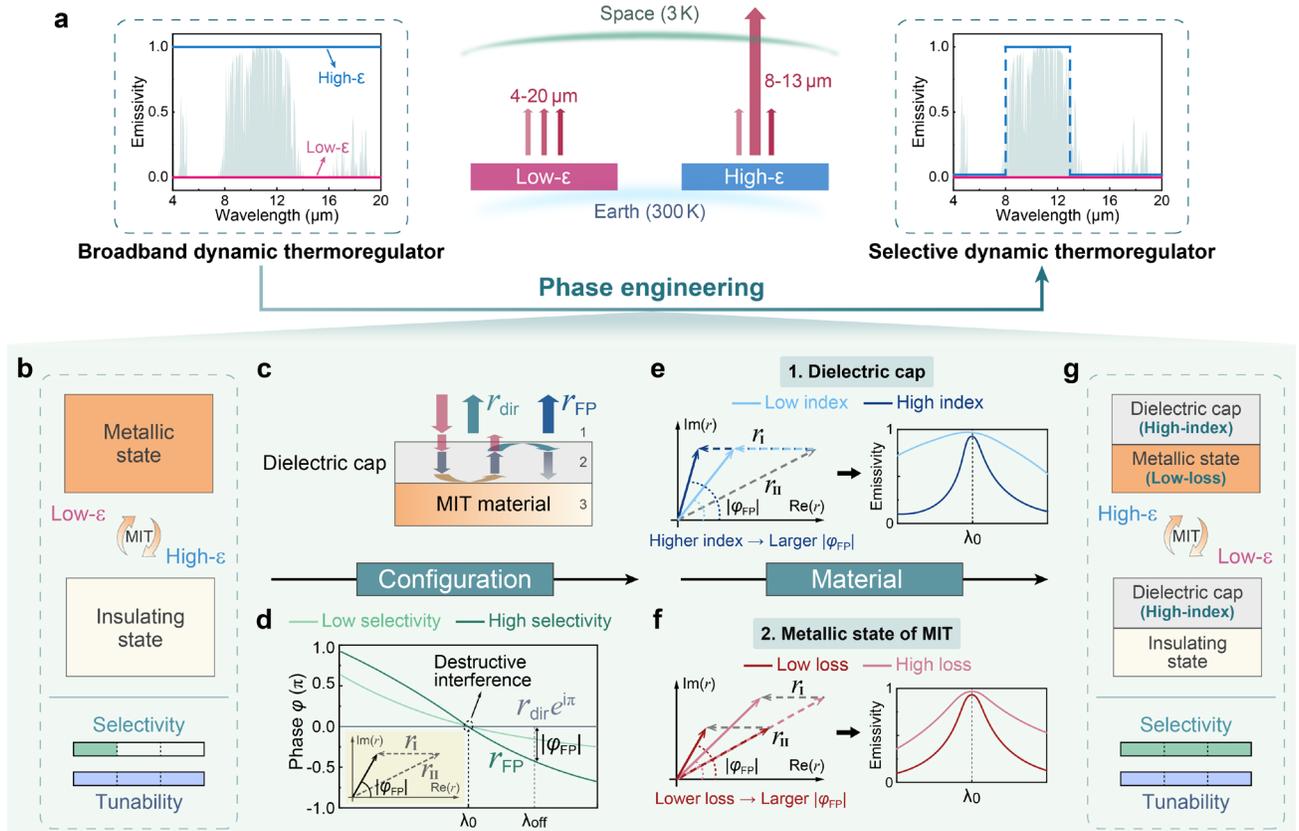

**Figure 1. Concept and implementation of spectrally-selective dynamic radiative thermoregulation via phase engineering.** (**a**) Comparison between ideal broadband (left) and ideal selective (right) dynamic thermoregulators. For the selective thermoregulator, its high-$\varepsilon$ state exhibits strong thermal emission only within the ATW of 8–13 μm. (**b**) Conventional MIT-based broadband dynamic emitter featuring high tunability but poor spectral selectivity. (**c**) Configuration of the selective thermoregulator, where a dielectric cap is placed atop the MIT material. The total reflection is decomposed into direct reflection $r_{\text{dir}}$ and FP reflection $r_{\text{FP}}$. (**d**) Schematic wavelength-dependent phase profiles of $r_{\text{dir}}e^{i\pi}$ and $r_{\text{FP}}$. Their intersection at $\lambda_0$ indicates a π-phase difference between $r_{\text{dir}}$

and $r_{FP}$, giving rise to destructive interference. At an off-resonant wavelength $\lambda_{off}$, a larger $|\varphi_{FP}|$ (i.e., larger phase difference $|\Delta\varphi|$) corresponds to higher spectral selectivity by causing a faster decay of destructive interference. Inset: phasor diagram of $r_I$ and $r_{II}$ that determine $|\varphi_{FP}|$. (**e**) Schematic phasor diagram showing how $|\varphi_{FP}|$ at $\lambda_{off}$ varies with dielectric-cap index, and the corresponding emissivity profiles. (**f**) Schematic phasor diagram showing how $|\varphi_{FP}|$ at $\lambda_{off}$ varies with the metallic-state loss of the MIT material, and the corresponding emissivity profiles. (**g**) Final design of a dynamic thermal emitter that exhibits both high spectral selectivity and tunability, achieved by placing a high-index dielectric cap on a MIT material which features a low-loss metallic state.

**Identification of material properties for high selectivity and tunability via Bayesian optimization**

To validate the theoretical analysis, we employed Bayesian optimization to globally search for the optimal material properties that enable high spectral selectivity and tunability (Fig. 2a). The optimization parameters include the refractive index $n$, extinction coefficient $k$, and thickness $t$ of the dielectric cap, as well as the Drude model parameters (plasma frequency $\omega_p$ and scattering frequency $\gamma$) which describe the complex permittivity $\tilde{\epsilon}$ of the MIT material in its metallic state. For the optimization objective, we define a figure of merit (FOM) as the product of spectral selectivity ($\xi$) and average emissivity tunability within 8–13 μm ($\Delta\tilde{\varepsilon}_{8-13\,\mu m}$) (top panel of Fig. 2a). The spectral selectivity is given by $\xi = \frac{\tilde{\varepsilon}_{8-13\,\mu m}}{\tilde{\varepsilon}_{4-8,13-20\,\mu m}} - 1$, where $\tilde{\varepsilon}_{8-13\,\mu m}$ and $\tilde{\varepsilon}_{4-8,13-20\,\mu m}$ are the 8–13 μm average emissivity within the corresponding wavelength ranges. The average emissivity tunability is defined as $\Delta\tilde{\varepsilon}_{8-13\,\mu m} = \tilde{\varepsilon}_{met} - \tilde{\varepsilon}_{ins}$, where $\tilde{\varepsilon}_{met}$ and $\tilde{\varepsilon}_{ins}$ are the average emissivity in the metallic and insulating states of the MIT material, respectively. We ran the Bayesian optimization algorithm for 3000 trials to ensure convergence toward the global maximum of FOM (Fig. S9). The optimization history of FOM with respect to the dielectric cap index $n$ and the metallic loss of the MIT material $\gamma/\omega_p$ is shown in the lower panel of Fig. 2a, clearly illustrating that higher $n$ and lower $\gamma/\omega_p$ are favorable for improved performance. This global optimization across the full parameter space confirms the key material property requirements for achieving high spectral selectivity and tunability, providing direct validation of our theoretical predictions.

In order to further confirm the critical role of a high dielectric index, we fix $n$ at different values while optimizing all other parameters. The resulting emissivity spectra are plotted in Fig. 2b, with the corresponding spectral selectivity and tunability summarized in Fig. 2c. For a low-index case (e.g., $n = 2$), both the spectral selectivity and emissivity contrast remain modest, even when the remaining parameters are optimized. The performance metrics are significantly improved by increasing $n$, highlighting the importance and effectiveness of a high-index dielectric cap in enhancing spectral control. Notably, the spectral selectivity continues to improve with increasing index, whereas the improvement of emissivity contrast slows down beyond $n = 4$. This is because the emissivity contrast is averaged over the 8–13 μm ATW range, and the over-enhanced selectivity suppresses the emission near 8 and 13 μm, thereby reducing the overall contrast across the target band (Fig. 2b). In other words, the trade-off between emissivity contrast and spectral selectivity is arbitrary and application-dependent. In Fig. 2c, we also present optimized results based on the optical properties of several representative materials, including AZ nLOF 2020 (nLOF) photoresist, zinc sulfide (ZnS), silicon (Si), Ge, and germanium antimony telluride ($Ge_2Sb_2Te_5$, or GST-225) (Figs. S10 and S11). These examples demonstrate that the observed trend is consistent across practical material systems, underscoring the feasibility of experimentally achieving such high performance.

Following the index analysis, we next examine the effect of the MIT material's metallic-state loss by fixing it at different levels and optimizing the other parameters. As shown in Fig. 2d, a high-loss metallic state yields broadband thermal emission, despite the optimization of other parameters including the dielectric-cap index. Figure 2e shows that reducing metallic loss substantially enhances the spectral selectivity. Under low-loss conditions, a slight decrease in emissivity contrast is observed, indicating an intrinsic trade-off associated with higher selectivity. We further perform optimization using the optical properties of representative MIT materials, including $VO_2$, electrochemically tunable polyaniline (PANI), and reversible metal electrodeposition (RME) systems with different working metals (Cu, Zn, and Ag) (Figs. S12 and S13). Among these, the RME systems, capable of reversibly depositing/dissolving metals from/into the electrolyte, exhibit the greatest potential to simultaneously achieve high spectral selectivity and tunability, owing to their exceptionally low metallic-state loss (Fig. 2e). These results highlight RME as a particularly promising route for experimentally realizing spectrally selective thermoregulators.

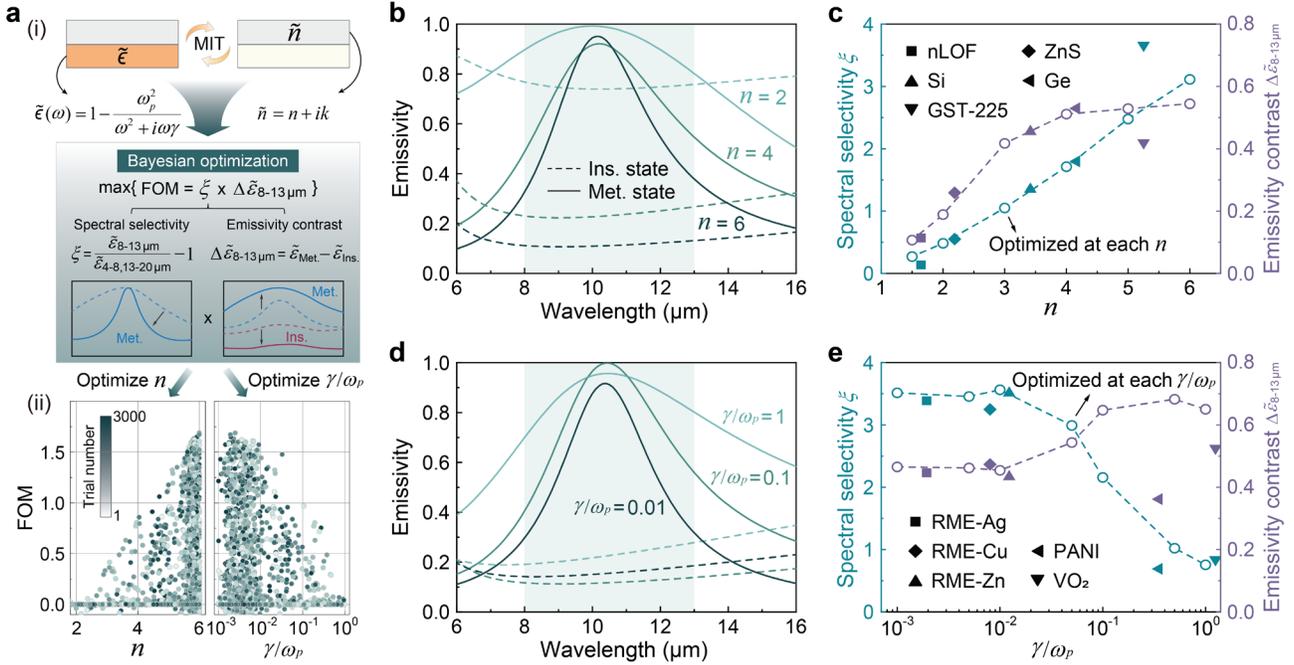

**Figure 2. Bayesian optimization for evaluating material-dependent performance in dynamic radiative thermoregulators.** (**a**) Schematic of the Bayesian optimization framework (i) and distribution of FOM as a function of $n$ and $\gamma/\omega_p$ during the optimization process (ii). Unless otherwise specified, the optimization ranges are: $n$ from 1 to 6, $k$ from $0.001n$ to $n$, $t$ from 1 nm to 10 μm, $\omega_p$ from $10^{14}$ to $10^{16}$ rad/s, and $\gamma$ from $0.001\omega_p$ to $\omega_p$. Without loss of generality, we fix the refractive index of the insulating state of the MIT material at $\tilde{n} = 1 + 0.01i$. (**b**) Optimized tunable emissivity spectra for cases with fixed dielectric-cap index $n = 2, 4$, and 6. (**c**) Optimized spectral selectivity and emissivity contrast as a function of $n$ (open markers). Solid markers show the results for representative dielectric materials, where the plotted $n$ values correspond to their refractive indices averaged over the 8–13 μm range. (**d**) Optimized tunable emissivity spectra for cases with fixed metallic-state losses $\gamma/\omega_p = 0.01, 0.1$, and 1. (**e**) Optimized spectral selectivity and emissivity contrast as a function of $\gamma/\omega_p$ (open markers). Solid markers are the results for representative MIT systems, with RME systems exhibiting the greatest potential for their lowest metallic-state losses.

**Highly-tunable spectrally-selective thermoregulator via reversible metal electrodeposition**

Building on the theoretical guidance established above, we proceeded to experimentally realize a dynamic thermal emitter that simultaneously achieves high spectral selectivity and tunability. For the MIT material, we employed the Cu-based RME system with an aqueous electrolyte, given its strong performance potential revealed in Fig. 2e. Cu was selected as the working metal primarily due to its favorable electrochemical properties, which ensure stable, durable, and rapid switching, as previously demonstrated in smart windows and dynamic radiative thermoregulation (*32*, *43*, *44*). In this RME system, Cu deposition corresponds to the metallic state, and its dissolution makes the electrolyte (assumed to be water in the calculations) serve as the insulating state. With RME-Cu identified as the MIT material, we optimized the dielectric cap properties to maximize the FOM of this system. The optimization reveals that, for the RME-Cu system, a lossy dielectric cap with $k \approx 0.3$ is required to achieve a high FOM (left side of Fig. 3a). This requirement arises because, in the metallic state, Cu itself has negligible optical loss and hence a dielectric cap with moderate loss is needed to provide sufficient absorption and enable strong thermal emission (see Fig. S14c versus Fig. S14d). Therefore, the RME-Cu system necessitates a high-index dielectric cap with an appropriate level of optical loss.

To relax the strict requirement for the dielectric-cap loss and use more real-world materials to accomplish high-FOM spectrally selective dynamic thermal emitters, we propose adding an ultrathin lossy coating on top of the dielectric layer to provide the desired optical absorption (right side of Fig. 3a). This configuration maintains spectral selectivity while enabling high emissivity tunability across a broader range of $k$ values, compared to the no-coating counterpart (Fig. S15). As a result, the coating not only boosts the FOM but also greatly broadens the range of compatible dielectric-cap materials (Figs. 3a and 3b). Common materials such as titanium (Ti), chromium (Cr), nickel (Ni), and ITO are promising candidates for the coating (Figs. S16 and S17). For example, the Ti coating enables optimal FOMs comparable to those obtained from an ideal coating material which is numerically optimized based on Drude model (Fig. 3b). Figure 3c further shows that the Ti coating significantly enhances the high-$\varepsilon$ state emissivity in the metallic MIT state, owing to the strong FP resonance that amplifies optical absorption within the ultrathin Ti film (4 nm). In this case, Ti coating accounts for 86% of the emissivity contribution, as highlighted in the inset of Fig. 3c. Importantly, this ultrathin Ti coating has a negligible impact on the emissivity in the low-$\varepsilon$ state when the strong FP resonance is absent, thus ensuring substantial emissivity tunability. Therefore, introducing such a lossy coating greatly facilitates experimental implementation of RME-Cu-based thermoregulators by easing the material constraints and expanding the palette of compatible dielectrics. The addition of an ultrathin coating can be viewed as a modification to the FP cavity top interface. Accordingly, the aforementioned phase analysis and theoretical insights remain valid for this updated configuration (see Supplementary Text 4). Nevertheless, we emphasize that the coating is not necessary if suitable lossy dielectric materials (e.g., GST-225) or RME-compatible lossy metals (e.g., Ti) are available, as illustrated in Fig. S14.

With the coating identified, we next investigated experimentally viable dielectric-cap materials for the RME-Cu device. We measured the mid-IR refractive indices of three candidates, nLOF, Si, and Ge, and computationally evaluated their potential in the Ti/dielectric/RME-Cu configuration. As expected, Ge yields the strongest spectral selectivity and emissivity contrast due to its highest refractive index among all candidates (Fig. S18). Experimentally, we fabricated three samples by integrating nLOF, Si, and Ge onto Cu substrates, followed by the deposition of a Ti coating. The emissivity measurements revealed a clear enhancement in spectral selectivity with increasing refractive index among the three materials ($n_{\text{nLOF}} < n_{\text{Si}} < n_{\text{Ge}}$) (Fig. 3d). Furthermore, this index-related spectral selectivity applies to almost all emission angles, as demonstrated in Fig. 3e, where the Ge case exhibits a much more suppressed emissivity at an off-resonant wavelength (e.g.,16 μm) than the nLOF counterpart. It is noteworthy that a higher index also helps suppress the angular dispersion of the interference wavelength, thereby maintaining the spectral profile over a wider range of angles (Fig. S19). For the Ge case, the spectral emissivity profile remains well preserved for emission angles from 0° to 80°, with the peak wavelength shifting by only 3% (Fig. 3f). This angular robustness is essential for radiative cooling applications, as it allows thermal radiation to consistently dissipate through the ATW.

To further validate the proposed design, we fabricated a prototype device based on the coating/dielectric/RME-Cu configuration, as schematically illustrated in Fig. 3g. We selected ZnSe as the superstrate for its high transmittance across a broad mid-IR range. On this superstrate, we deposited 7 nm Ti as the coating and subsequently 565 nm Ge to form the cavity layer. To enable the RME of Cu, a monolayer graphene was transferred to the Ge as the working electrode. The aqueous electrolyte mainly consisted of $Cu(ClO_4)_2$, and a Cu foil was used as the counter electrode. Upon applying a cathodic potential to the graphene working electrode, $Cu^{2+}$ cations were electrochemically reduced and deposited into a continuous Cu thin film, as characterized by the cross-sectional scanning electron microscopy (SEM) (Fig. 3h).

The thermal radiation tunability was first characterized by IR camera imaging. As shown in Fig. 3i, when the device was placed on a hot plate at 60 °C, a substantial surface temperature difference (>20 °C) between the low-$\varepsilon$ and high-$\varepsilon$ states was uniformly observed across the device area (~2×2 cm$^2$). More importantly, the high-$\varepsilon$ state exhibited great spectral selectivity within the ATW, as demonstrated by the measured emissivity spectra in Fig. 3j. Note that the slight discrepancy between experimental and simulated spectra is attributed to the influence of ZnSe superstrate and its antireflection coating, which are not included in simulation due to the unavailability of their properties from the supplier. Through Cu RME, the active device experimentally achieved a substantial peak emissivity modulation of 0.76 at 9.9 μm and an average tunability of 0.57 over the 8–13 μm range. These results experimentally validate the efficacy of our theoretical analysis and device design principles in realizing active thermal emitters that combine substantial emission tunability with spectral selectivity.

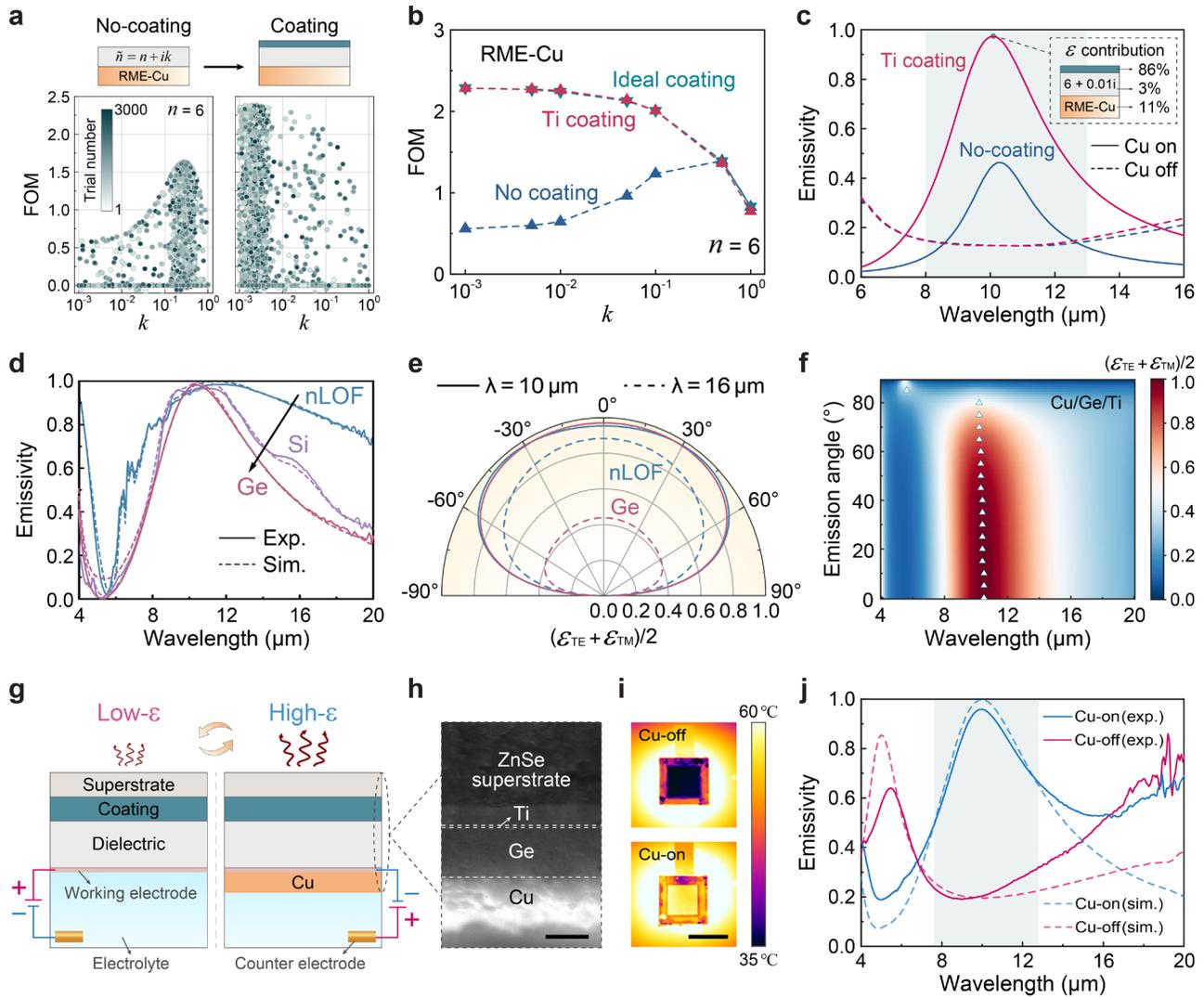

**Figure 3. Dynamic radiative thermoregulator with high spectral selectivity and tunability enabled by reversible Cu electrodeposition.** (**a**) Performance improvement through the addition of a lossy coating in the RME-Cu device. The bottom panels show the distribution of FOM as a function of the dielectric-cap extinction coefficient $k$ during Bayesian optimization for configurations without (left) and with (right) the coating. (**b**) Optimized FOMs for RME-Cu devices under different configurations, with dielectric-cap $k$ fixed at various values during optimization. (**c**) Simulated emissivity spectra of RME-Cu devices with and without a Ti coating. The dielectric-cap index is set to $n = 6 + 0.01i$, and the thicknesses of the coating and dielectric layers are optimized to maximize

FOM. Inset: layer-resolved contributions to emissivity (absorptivity) at 10 μm for the Ti-coating case. (**d**) Simulated and measured emissivity spectra of Ti/dielectric/Cu structures using nLOF, Si, and Ge as dielectric materials. The arrow indicates the direction of increasing refractive index across the three materials. Structural parameters of these configurations are provided in Fig. S19. (**e**) Polar plot of simulated emissivity for the nLOF and Ge cases at 10 and 16 μm, corresponding to on- and off-resonant wavelengths, respectively. Emissivity is averaged over transverse electric (TE) and transverse magnetic (TM) polarizations. (**f**) Simulated angle-resolved emissivity map of the Ti/Ge/Cu configuration, with triangles marking the spectral peak positions every 5° from 0° to 85°. (**g**) Schematic of the coating/dielectric/RME-Cu device, capable of dynamically switching between high-$\varepsilon$ and low-$\varepsilon$ states through reversible Cu electrodeposition on the working electrode. (**h**) Cross-sectional SEM image of the device using Ti and Ge as coating and dielectric layers, respectively, with ~200 nm of electrodeposited Cu. Scale bar, 500 nm. (**i**) IR camera images of the device on a 60 °C hot plate, showing the low-$\varepsilon$ (Cu off, top) and high-$\varepsilon$ (Cu on, bottom) states. Scale bar, 2 cm. (**j**) Simulated and measured emissivity spectra of the device in the Cu-on (high-$\varepsilon$) and Cu-off (low-$\varepsilon$) states.

**Demonstration of multispectral electrochromic window**

Beyond mid-IR tuning, controlling the spectral response in the solar spectrum is also crucial for effective thermoregulation in many applications. A representative case is smart windows, where the mid-IR emissivity dictates the radiative heat dissipation and the solar transmittance governs the indoor heating. Many MIT materials, such as conjugated polymers and RME systems, can offer substantial optical tunability spanning from visible to mid-IR, and have been widely explored for electrochromic smart windows (*45*, *46*). However, their multispectral responses are often not ideal or even contradictory in the context of thermoregulation. As illustrated in Fig. 4a, the metallic state blocks solar radiation and facilitates cooling, but also exhibits low mid-IR emissivity that promotes heating. On the other hand, the insulating state transmits solar radiation to support indoor heating, but also suffers from substantial radiative heat loss due to its high emissivity. To reconcile these conflicting responses, a dielectric cap can be introduced atop the MIT material to reshape its mid-IR response, thereby harmonizing solar and thermal management while preserving spectral tunability (Fig. 4b). This structural modification enables desirable multispectral electrochromism, allowing dynamic switching between solar heating and passive radiative cooling in response to environmental conditions.

The realization of multispectral electrochromic windows (Multi-ECWs) demands a dielectric-cap material that combines high refractive index in the mid-IR range with good transparency in the visible. However, this is practically challenging as many mid-IR high-index materials, such as Si and Ge, are opaque in the visible, while most visibly transparent materials like polymers and dielectrics (e.g., silicon dioxide and aluminum oxide) exhibit low indices or undesirable optical resonances in the mid-IR. To address this challenge, we developed a composite material by embedding ITO NCs into a broadband transparent polymer matrix of nLOF. In the visible regime, both ITO and nLOF are transparent for their wide bandgaps (*47*). In mid-IR, the nanometer size of ITO NCs (~20 nm), being three orders of magnitude smaller than the wavelengths (~10 μm), enables efficient light transmission. As a result, the ITO NCs behave as a dielectric medium in the mid-IR, with optical properties dictated by the effective medium theory. By optimizing the filling ratio (Fig. S21) and doping level of ITO NCs (Fig. S22), the effective mid-IR index can be engineered to as high as ~3 while maintaining high visible transparency, showing great potential to enable the desired multispectral electrochromism.

Figure 4c shows the schematic of a Multi-ECW design based on the Cu RME. For device fabrication, we used barium fluoride ($BaF_2$) superstrate for its high transparency from visible to mid-IR. A 15-nm-thick ITO film was sputtered onto the substrate to serve as the lossy coating. To construct the dielectric cap, we alternately spin-coated ITO NCs and nLOF in a layer-by-layer manner until the device exhibited a mid-IR reflectivity peak near 10 μm, indicating that the desired FP cavity thickness had been achieved (Fig. S24). At this thickness (~1 μm), Cu electrodeposition flips the reflectivity peak into a dip (i.e., an emissivity peak) within the ATW, by introducing a π-phase shift at the bottom cavity interface which turns constructive interference of reflection waves into destructive interference. To perform Cu RME on the dielectric cap, we used graphene coated with platinum (Pt) as the working electrode for its wide-band transparency and ability to support uniform Cu deposition (*32*, *35*, *48*). By applying a negative voltage (e.g., -1.5 V) against a Cu-foil counter electrode, a continuous and uniform Cu layer can be deposited on the Pt-modified graphene electrode, as shown in the cross-sectional SEM images in Fig. 4d and Fig. S25. In this way, the RME process enables the device to switch between a visibly transparent, low-$\varepsilon$ mode for heating and a visibly opaque, high-$\varepsilon$ mode for cooling.

To proceed with the experimental demonstration, we measured the refractive indices nLOF with and without ITO NCs filling. As shown in Fig. 4e, incorporation of ITO NCs substantially increased the polymer index by 64%, reaching $n \approx 2.7$ within the ATW range of 8–13 μm. This high-index dielectric cap enables strong mid-IR emission with considerable spectral selectivity at the ATW when the Cu is deposited, leading to a desirable opaque and high-$\varepsilon$ state (Fig. 4f). When Cu is dissolved, the device returns to a transparent state with an average visible transmittance of 65% and reduced mid-IR emissivity. The device demonstrated a peak emissivity tunability of 0.28 at ~10 μm and an average of 0.19 across the ATW range. We note that further enhancement of emissivity contrast could be achieved by engineering the dielectric-cap material to higher index, for instance by increasing the ITO NCs filling ratio. Figure 4g shows the fabricated device on a 2.5 × 2.5 cm² $BaF_2$ superstrate, which exhibits uniform Cu electrodeposition and stripping during switching, as characterized by the visible and IR images in Fig. 4h. In the experiment, we applied -1.5 V for 10 s to electrodeposit Cu, followed by 0.5 V for 2 min for a complete stripping. The device's cyclability was evaluated for 120 switching cycles, with IR camera–measured surface temperatures confirming stable and reversible performance throughout the test (Fig. 4i).

With its ability to simultaneously modulate solar and mid-IR spectra, the proposed Multi-ECW holds great promise for building thermal regulation, particularly in urban areas where spectrally selective radiative cooling is essential during hot weather (*18*, *19*) (Fig. 4j). To evaluate its energy-saving ability, we used EnergyPlus, a building energy simulation software developed by the U.S. Department of Energy (DOE), to model the year-round thermoregulation energy consumption of buildings equipped with the Multi-ECW. Its tunable solar transmittance and mid-IR emissivity (Fig. 4f) were imported as window properties (Table S2). The simulations were conducted using the DOE new-2004 midrise apartment model with a 15% window-to-wall ratio (Fig. S26), across 15 U.S. cities each representing a distinct climate zone (Fig. S27). The results show that the Multi-ECW achieves considerable energy savings across all climate zones, with an average saving of 15 MJ/m$^2$ and up to >30% reduction in total energy use in hot areas such as Los Angeles, compared to the commercial low-$\varepsilon$ glass (Fig. S28).

Given the consistently increasing window-to-wall ratios in modern architecture, the energy-saving potential of Multi-ECW systems is expected to become even more significant in the future.

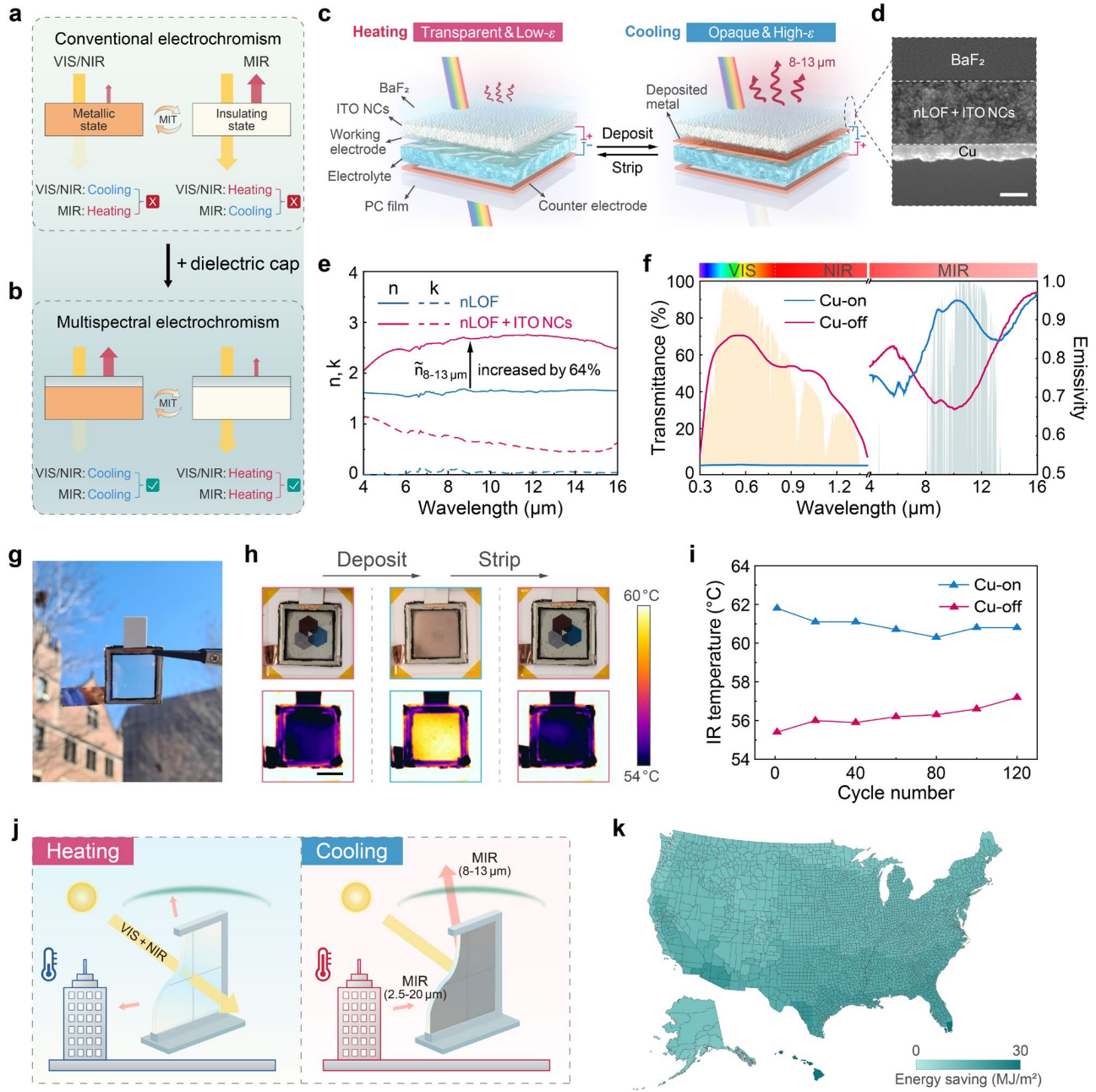

**Figure 4. Multispectral electrochromic window (Multi-ECW) for building thermoregulation. (a)** Conventional MIT-based electrochromic system exhibiting a functional conflict between solar and mid-IR responses for thermoregulation. **(b)** Multispectral electrochromism harmonizing solar and mid-IR responses, realized by introducing a dielectric cap to the MIT material. (**c**) Schematic of the Multi-ECW structure, electrically switchable between a transparent, low-$\varepsilon$ mode for solar heating and an opaque, spectrally selective high-$\varepsilon$ mode for radiative cooling. (**d**) Cross-sectional SEM image of the Multi-ECW device after Cu electrodeposition. (**e**) Enhancement of the mid-IR refractive index of the transparent polymer (i.e., nLOF) via incorporation of ITO NCs. (**f**) Measured visible (VIS) and near-IR (NIR) transmittance and mid-IR (MIR) emissivity spectra of the Multi-ECW in the Cu-on and Cu-off states. (**g**) Photograph of the fabricated Multi-ECW device. (**h**) Optical and IR images of the Multi-ECW device during Cu electrodeposition and stripping, demonstrating uniform and reversible switching. Scale bar, 1 cm. (**i**) Cycling stability of the Multi-ECW device, characterized by IR camera-measured apparent temperature. (**j**) Conceptual application of Multi-ECW applied to urban buildings, enabling dynamic switching between solar heating and spectrally selective radiative cooling in response to ambient temperature variations. (**k**) Simulated energy savings across the 15 U.S. climate zones when implementing Multi-ECWs in buildings.

**Discussion**

We developed a design framework for spectrally selective dynamic thermal emitters by introducing a dielectric cap onto MIT materials. The dielectric cap acts as an asymmetric FP cavity that enables spectral engineering beyond the intrinsic optical response of MIT systems. Our phasor diagram analysis and Bayesian optimization revealed that the spectral selectivity is dictated jointly by the dielectric-cap index and the metallic-state loss of the MIT material. Guided by these insights, we experimentally achieved a spectrally selective dynamic emitter using Ge as the dielectric cap and Cu RME as the active MIT material, capable of electrically tuning the thermal emittance in the ATW from

~0.2 to ~0.9. We further extended this dielectric capping approach to resolve the functional conflict between solar heating and radiative cooling inherent to conventional MIT-based electrochromic systems, enabling the desired multispectral electrochromism with synergized solar and mid-IR responses for efficient thermoregulation. By developing a transparent high-index composite dielectric based on ITO NCs, we demonstrated a multispectral electrochromic system with desirable visible, near-IR, and mid-IR tunability, showcasing substantial energy-saving potential for thermoregulation.

Given its simplicity and adaptability, the dielectric capping strategy can be applied across a wide range of MIT platforms, regardless of their physical forms, switching stimuli, or working principles. Its flexible control over spectral width and target wavelength range further enables versatile tailoring of devices for different thermal management needs. Collectively, these findings establish a general design paradigm for dynamic spectral control, with broad implications for advanced thermoregulation and beyond.

## ACKNOWLEDGEMENTS

This work used the shared facilities in the Searle Cleanroom and the Pritzker Nanofabrication Facility, part of the Pritzker School of Molecular Engineering at the University of Chicago, supported by Soft and Hybrid Nanotechnology Experimental (SHyNE) Resource (NSF-ECCS-2025633), a node of the National Science Foundation's National Nanotechnology Coordinated Infrastructure (RRID: SCR_022955).

**Funding:** The project was supported by the startup fund from the Pritzker School of Molecular Engineering at the University of Chicago and the National Science Foundation (Electrical, Communications and Cyber Systems award no. 2324286).

**Author contributions**: P.-C.H. and Q.L. conceived the idea. Q.L. performed the theoretical analysis, device design, simulations, fabrications, electrochemical experiments, and characterizations. Y.C. conducted the ITO NCs synthesis, ITO NCs properties characterization, and helped in Multi-ECW device fabrication and characterization. Z. L. set up the Bayesian optimization algorithm. C.S. assisted with energy consumption calculation. X.W. carried out the IR permittivity measurements. R.W., Q.F., C.-T.F., and P.-J.H. assisted with UV–Vis measurements, ITO NC synthesis, graphene transfer, and IR camera imaging, respectively. G.Y., G.H., and T.-H.C. contributed to sample preparation.

## CONFLICT OF INTEREST

The authors declare no conflict of interest.

5. X. Yin, R. Yang, G. Tan, S. Fan, Terrestrial radiative cooling: Using the cold universe as a renewable and sustainable energy source. *Science* **370**, 786–791 (2020).

6. Q. Cheng, G. W. Ho, A. P. Raman, R. Yang, Y. Yang, Regulating thermal radiation for energy and sustainability. *Next Energy* **1**, 100019 (2023).

7. J. Liang, J. Wu, J. Guo, H. Li, X. Zhou, S. Liang, C.-W. Qiu, G. Tao, Radiative cooling for passive thermal management towards sustainable carbon neutrality. *Natl. Sci. Rev.* **10**, nwac208 (2023).

8. A. P. Raman, M. A. Anoma, L. Zhu, E. Rephaeli, S. Fan, Passive radiative cooling below ambient air temperature under direct sunlight. *Nature* **515**, 540–544 (2014).

9. P. C. Hsu, A. Y. Song, P. B. Catrysse, C. Liu, Y. Peng, J. Xie, S. Fan, Y. Cui, Radiative human body cooling by nanoporous polyethylene textile. *Science* **353**, 1019–1023 (2016).

10. Y. Zhai, Y. Ma, S. N. David, D. Zhao, R. Lou, G. Tan, R. Yang, X. Yin, Scalable-manufactured randomized glass-polymer hybrid metamaterial for daytime radiative cooling. *Science* **355**, 1062–1066 (2017).

11. K. Lin, S. Chen, Y. Zeng, T. C. Ho, Y. Zhu, X. Wang, F. Liu, B. Huang, C. Y.-H. Chao, Z. Wang, C. Y. Tso, Hierarchically structured passive radiative cooling ceramic with high solar reflectivity. *Science* **382**, 691–697 (2023).

12. X. Zhao, T. Li, H. Xie, H. Liu, L. Wang, Y. Qu, S. C. Li, S. Liu, A. H. Brozena, Z. Yu, J. Srebric, L. Hu, A solution-processed radiative cooling glass. *Science* **382**, 684–691 (2023).

13. S. Fan, W. Li, Photonics and thermodynamics concepts in radiative cooling. *Nat. Photonics* **16**, 182–190 (2022).

14. K. Yang, X. Wu, L. Zhou, P. Wu, I. Gereige, Q. Gan, Towards practical applications of radiative cooling. *Nat. Rev. Clean Technol.* **1**, 235–254 (2025).

15. R. Liu, S. Wang, Z. Zhou, K. Zhang, G. Wang, C. Chen, Y. Long, Materials in Radiative Cooling Technologies. *Adv. Mater.* **37**, 2401577 (2025).

16. J. Zhou, T. G. Chen, Y. Tsurimaki, A. Hajj-Ahmad, L. Fan, Y. Peng, R. Xu, Y. Wu, S. Assawaworrarit, S. Fan, M. R. Cutkosky, Y. Cui, Angle-selective thermal emitter for directional radiative cooling and heating. *Joule* **7**, 2830–2844 (2023).
26